\documentclass[10pt,amssymb,superscriptaddress,showkeys,twocolumn,prl]{revtex4-2}
\usepackage{graphicx}
\usepackage{dcolumn}
\usepackage{bm}
\usepackage{changes}
\usepackage{amsmath,amssymb}
\makeatletter
\begin{document}
\title{Free Extension of Topological States via Double-zero-index Media}

\author{Rui Dong}
\thanks{These authors contributed equally to this work.}
\affiliation{Key Laboratory of State Manipulation and Advanced Materials in Provincial Universities, School of Physics and Technology, Nanjing Normal University, Nanjing 210023, China}
\author{Changhui Shen}
\thanks{These authors contributed equally to this work.}
\affiliation{MOE Key Laboratory of Modern Acoustics, National Laboratory of Solid State Microstructures, School of Physics, Collaborative Innovation Center of Advanced Microstructures, and Jiangsu Physical Science Research Center, Nanjing University, Nanjing 210093, China}
\author{Changqing Xu}
\email{changqing.xu@nnu.edu.cn}
\affiliation{Key Laboratory of State Manipulation and Advanced Materials in Provincial Universities, School of Physics and Technology, Nanjing Normal University, Nanjing 210023, China}
\author{Yun Lai}
\email{laiyun@nju.edu.cn}
\affiliation{MOE Key Laboratory of Modern Acoustics, National Laboratory of Solid State Microstructures, School of Physics, Collaborative Innovation Center of Advanced Microstructures, and Jiangsu Physical Science Research Center, Nanjing University, Nanjing 210093, China}
\author{Ce Shang}
\email{shangce@aircas.ac.cn}
\affiliation{Aerospace Information Research Institute, Chinese Academy of Sciences, Beijing 100094, China}

\begin{abstract}
Topological states, known for their robustness against disorder, offer promising avenues for disorder-resistant devices. However, their intrinsic spatial confinement at interfaces imposes geometric constraints that limit the scalability of topological functionalities. Here, we propose a strategy to overcome this limitation by using double-zero-index media to expand topological interfaces. Although occupying finite space, these media are optically equivalent to infinitesimal points, effectively altering the geometry of topological interfaces and breaking conventional bulk-edge correspondence. This strategy enables the spatial expansion of uniform topological states beyond their native interface, offering new possibilities for topological photonic devices. We have verified this behavior through numerical simulations and microwave experiments in a two-dimensional photonic Su–Schrieffer–Heeger lattice. Our findings offer a universal framework to overcome the inherent dimensional limitations of topological states, with implications extending to general wave systems such as acoustic metamaterials.

\end{abstract}
                              \maketitle

In recent decades, topological insulators have attracted widespread interest in artificial systems due to their unique bulk-edge correspondence \cite{PhysRevLett.49.405,RevModPhys.82.3045,RevModPhys.83.1057,PhysRevX.9.011037,PhysRevLett.134.206603,PhysRevLett.134.136601,sciadv.adq9285,PhysRevLett.132.233801}, which has been harnessed to manipulate classical waves in innovative ways and offered exciting opportunities for practical applications. For instance, photonic analogues of topological systems have demonstrated robust unidirectional edge states that are immune to defects and disorder \cite{nature12066,nature08293,khanikaev2013photonic,wu2015scheme,shalaev2019robust,dong2017valley,PhysRevLett.134.033803,10.1002/apxr.202200053,khanikaev2024topological}. Recent developments in higher-order topological states and disclination states have enabled novel light confinement mechanisms, providing the foundation for next-generation topological lasers distinguished by low thresholds and strong directionality \cite{harari2018topological,bandres2018topological,dikopoltsev2021topological,zhang2020low,PhysRevLett.133.133802,science.adr5234,PhysRevLett.133.233804,Liu2021bulk}. However, the inherent spatial localization of topological states at interfaces results in a highly restricted mode volume. As a result, phenomena such as Purcell enhancement  \cite{PhysRevLett.120.114301} are significantly suppressed. Moreover, their working wavelengths are typically comparable to the lattice constant \cite{zhang2020higher,jiang2024observation}. Thus, despite the achievement of high-quality factors, the spatial extent of topological functionalities is limited. This poses a major challenge for the scalability and integration of topological photonic devices, particularly those based on higher-order topological phases. Despite extensive research, strategies to overcome this fundamental limitation remain scarce.

\begin{figure}[t!]
\centering
\includegraphics[width=0.9\columnwidth]{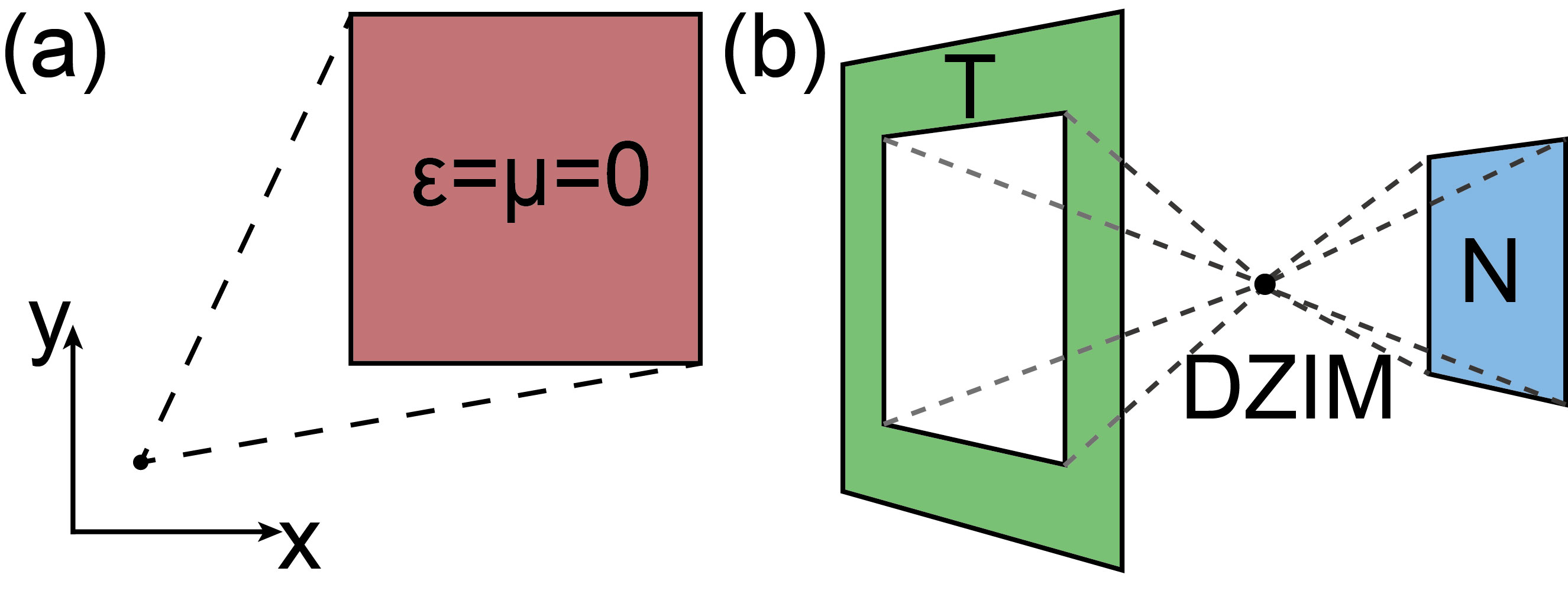}
\caption{Double-zero-index media and the reshaping of topological interfaces. (a) The expansion of an infinitesimal spatial point into a square of DZIM. (b) In optical space, topological trivial (labeled ``T") and non-trivial (labeled ``N") bulk materials are effectively connected by a 0D interface when the DZIM is inserted between them.}
\label{fig1}
\end{figure}

Double-zero-index media (DZIM), characterized by simultaneously near-zero permittivity and permeability, can be comprehended by isotropically expanding a zero-dimensional (0D) point into a finite region with the principles of transformation optics as shown in Fig.~\ref{fig1}(a) \cite{xu2021three,Supp}. Although it occupies a finite space, such a medium is optically equivalent to the original point. Thus, a plane wave passing through this space does not accumulate any phase. In a two-dimensional (2D) electromagnetic DZIM, Maxwell’s equations force the electric (or magnetic) field to be uniform due to the vanishing curl, reflecting its equivalence with a 0D point. Thanks to this zero-index nature, DZIM has attracted significant interest, such as wavefront tailing \cite{liberal2017near,alu2007epsilon,moitra2013realization,li2015chip}, impurity cloaking \cite{huang2011dirac,luo2015unusual}, wave tunneling \cite{fleury2013extraordinary,edwards2008experimental,PhysRevLett.124.074501,vzp2-6gns}, photonic doping \cite{liberal2017photonic,liberal2016nonradiating,hao2010super,li2022geometry,li2022performing} and “anti-doping” effect \cite{xu2021three}. Very recently, a unique bulk-spatiotemporal vortex correspondence, different from conventional bulk-edge correspondence, bridges DZIM and topological photonics \cite{zhang2025bulk}.

\begin{figure*}[t!]
\centering
\includegraphics[width=1.5\columnwidth]{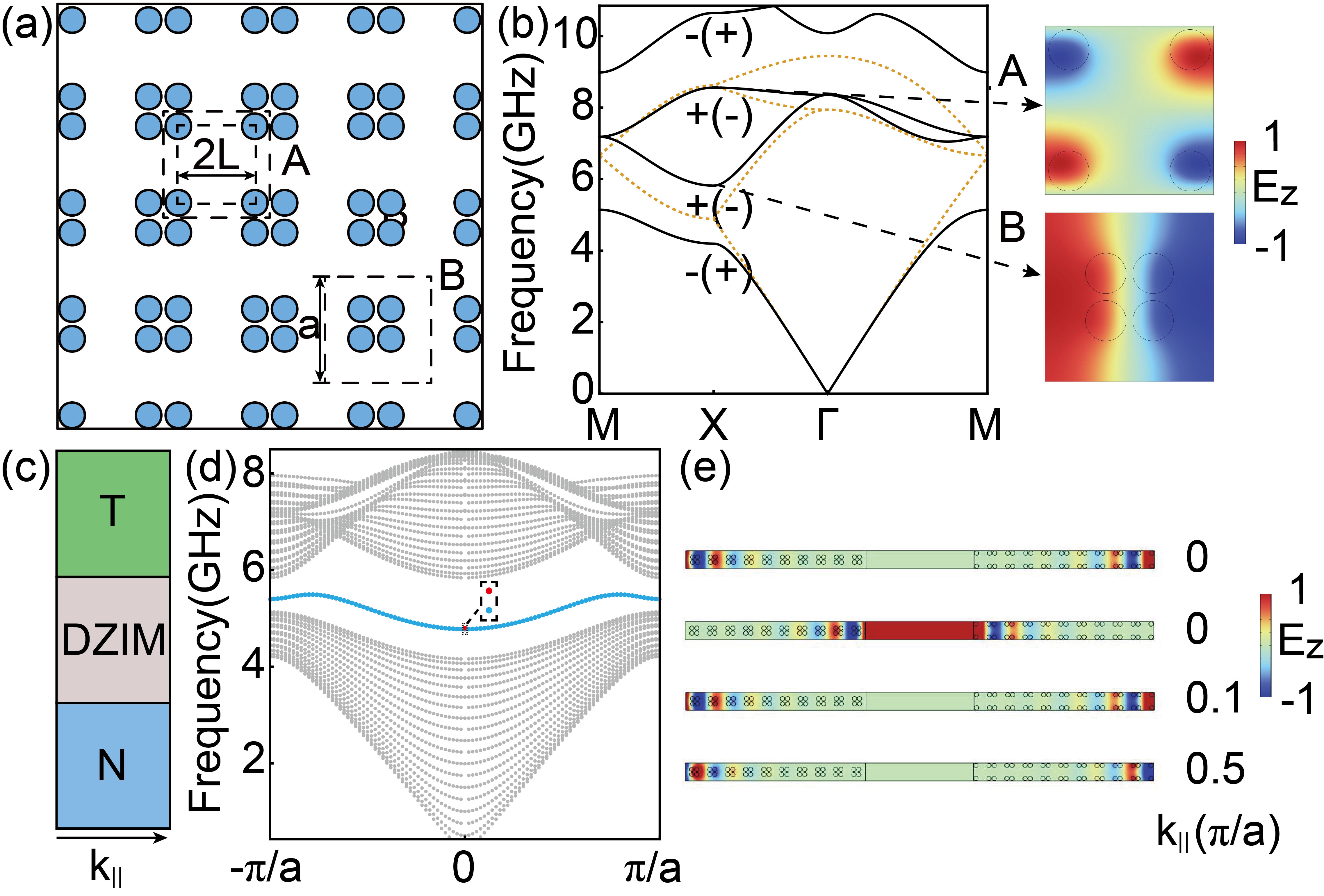}
\caption{A topological edge state extended by the DZIM layer. (a) A PC with different unit cell selections, labeled A and B. Each unit cell is composed of 4 cylinders with permittivity $\varepsilon=7.7$ and radius $r=2.4~\mathrm{mm}$. The lattice constant is $a=20~\mathrm{mm}$. Each rod is displaced by a distance $L$ from the unit cell center in both $x-$  and $y-$directions. (b) Black solid lines: the calculated band structures for $L=7.2(2.8)~\mathrm{mm}$. The yellow dotted lines: the calculated band structures for $L=5~\mathrm{mm}$. The symbols ``$\pm$"  represent the parity of the states under $\pi$ rotation, obtained from the eigenfields. The right insets are illustrative examples. The cases of $L=7.2~\mathrm{mm}$ and $L=2.8~\mathrm{mm}$ share the same band structure but have different bulk polarizations. (c) Scheme for a DZIM layer sandwiched by the topological trivial PC (labeled ``T") and the nontrivial PC (labeled ``N"). The upper/lower and left/right boundaries of the three-layered structure are two pairs of periodic boundary conditions. (d) The projected band structure along the direction of $k_\parallel$ in (c) exhibits bulk states (gray), 1D topological edge states (blue) at the upper/lower interface, and the spatially extended edge state (red) that spans across the DZIM layer. (e) The electric field distributions for the edge states when $k_\parallel=0$, $k_\parallel=0.1\pi/a$, and $k_\parallel=0.5\pi/a$, respectively. }
\label{fig2}
\end{figure*}

In this work, we propose a strategy to overcome the intrinsic spatial localization of topological states by separating topological interfaces and filling the gap with DZIM. We consider a 2D Su–Schrieffer–Heeger (SSH) model implemented in a dielectric photonic crystal (PC). By tuning a geometric parameter in each unit cell, the PC undergoes a topological transition between configurations with different quantized bulk polarizations, resulting in the emergence of one-dimensional (1D) edge states and 0D corner states. As illustrated in Fig.~\ref{fig1}(b), when we insert the DZIM between topologically trivial and non-trivial bulk materials, they behave as if they were connected by a 0D interface in optical space. Thus, the in-phase topological states are spatially extended, while the out-of-phase topological states are suppressed. Due to its equivalence to a point, the DZIM effectively reshapes the geometry of the topological interface, thus modifying the manifestation of bulk–edge correspondence. Nevertheless, distinct bulk polarizations across the interface remain essential for the emergence of topological states, ensuring that their topological origin is preserved. This approach is validated through numerical simulations and microwave experiments. Our findings demonstrate that DZIM enables the spatial extension of topological states beyond their interfaces while retaining their topological characteristics. This strategy may provide a foundation for photonic platforms that benefit from enhanced spatial scalability and mode delocalization.


As shown in Fig.~\ref{fig2}(a), we consider a 2D PC consisting of a square lattice of dielectric rods with relative permittivity $\varepsilon=7.7$, radius $r=2.4 ~\mathrm{mm}$, and lattice constant $a=20 ~\mathrm{mm}$. A geometric parameter $L$ defines the displacement of each rod from the center of the unit cell in both the horizontal and vertical directions. The dashed boxes in Fig.~\ref{fig2}(a) are two unit cells: configuration A with $L=7.2 ~\mathrm{mm}$ and configuration B with $L=2.8~\mathrm{mm}$, with reversed choices of center and corner, respectively. Along high symmetry lines of the irreducible Brillouin zone, the band structure for transverse electric ($E_z$) polarized waves is calculated by COMSOL Multiphysics and plotted in Fig.~\ref{fig2}(b). Although the eigenfrequencies are invariant under different unit cell choices, the corresponding eigenfields exhibit distinct spatial symmetries, as illustrated by the insets. The symbols ``+" and ``-" indicate the eigenvalues ``+1" and ``-1" under $\pi$ rotation for the $i_{\text{th}} (i=1,2,3,4)$ eigenstates at the $X$ point.

The PC can be described by a 2D SSH model, which undergoes a topological phase transition when $L$ crosses $a/4$ \cite{zhang2019dimensional,chen2019corner}. The topological phase of the PC can be characterized by a 2D bulk polarization $P$ in terms of the parities of eigenstates at the high-symmetry points:
\begin{equation}
P_n=\frac{1}{2}(\sum\limits_iq^i_n {\text{mod}} 2),\quad (-1)^{q^i_n}=\frac{\eta(X_n^i)}{\eta(\Gamma^i)},
\end{equation}
where the summation is taken over all the occupied bands, $\eta$ denotes the parity associated with $\pi$ rotation, and $n$ stands for $x$ or $y$. Under the first band gap, we obtain quantized bulk polarizations $P = (1/2, 1/2)$ and $P = (0,0)$ for unit cell selections A and B, which indicates nontrivial and trivial topologies, respectively. At $L=5 ~\mathrm{mm}$, the two unit cell configurations merge and the bandgap closes, marking the topological phase transition. This transition point enables the double-zero-index property in wave manipulation, which will be demonstrated later.

When two PCs with distinct bulk polarizations are bonded together, a hierarchy of topological states emerges within the band gap \cite{zhang2019dimensional,zhang2019second}. To spatially extend these states, we introduce a DZIM layer at the interface. As illustrated in Fig.~\ref{fig2}(c), a DZIM slab is inserted between two PC bulks with trivial (green) and nontrivial (blue) topologies. The structure is bounded by periodic boundary conditions along both the vertical and horizontal directions. Along the direction of $k_\parallel$ in Fig.~\ref{fig2}(c), the projected band structure of this three-layered structure is shown in Fig.~\ref{fig2}(d). Before the insertion of the DZIM layer, each of the two interfaces between topologically distinct bulks supports edge states at the same frequency, consistent with the bulk-edge correspondence. Due to its vanishing effective permittivity and permeability, the DZIM behaves as a so-called photonic ``void space"\cite{xu2021three}, supports only uniform electromagnetic fields inside, and thus enables the extension of in-phase topological states. It arises from the curl-less nature of a 2D electromagnetic DZIM, which enforces a vanishing electric field gradient and thus a uniform field in DZIM. As a result, specifying the electric field at a single point effectively determines the field throughout the entire DZIM. At $k_\parallel=0$, the in-phase topological state (red dot) extends efficiently into the DZIM layer and maintains nearly the same frequency as the original interface state. However, when $k_\parallel \ne 0$, the out-of-phase topological interface states are incompatible with the uniform field condition enforced by the DZIM and thus cannot be supported at the interface with DZIM layer. This behavior is confirmed by the electric field distributions of the topological edge states shown in Fig. 2(e), where only the state at $k_\parallel = 0$ is extended through the DZIM layer.

\begin{figure}[htp!]
\centering
\includegraphics[width=1\columnwidth]{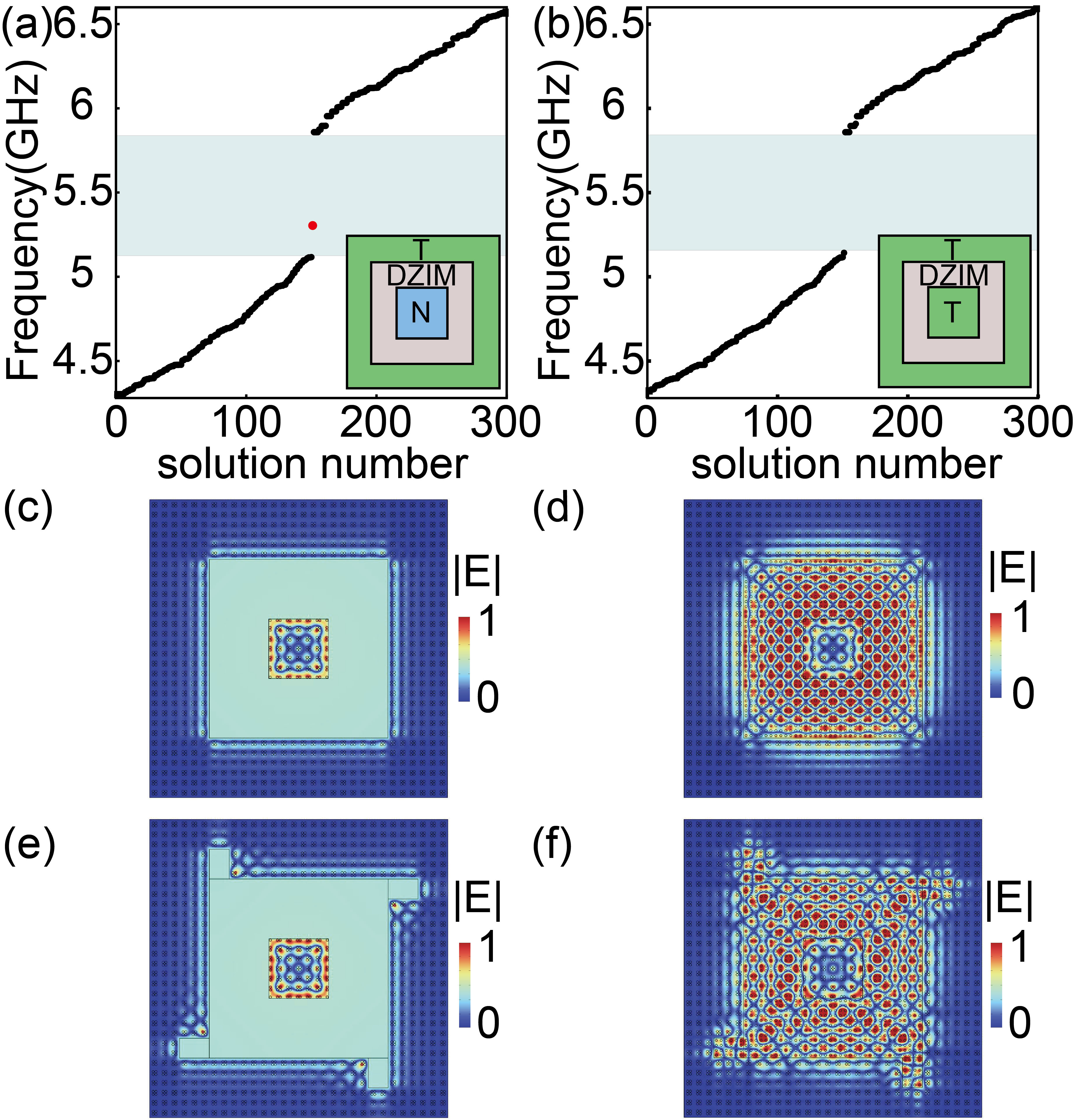}
\caption{A topological corner state extended by DZIM layer. (a) The eigenfrequency spectrum when the DZIM is embedded in a square interface between the topologically trivial and nontrivial bulks. (b) The eigenfrequency spectrum when the DZIM is embedded between two topologically trivial bulks. (c, d) The electric field distribution of the spatially extended corner state when the filling is set as (c) ideal DZIM and (d) a PC with the effective double-zero-index property, respectively. (e, f) The electric field distribution of the spatial extension of the corner state when the shape of (e) the ideal DZIM and (f) the PC are changed.}
\label{fig3}
\end{figure}

Furthermore, we apply the concept of spatial extension to 0D corner states by introducing a DZIM layer between two topologically distinct PC bulks. As illustrated in the inset of Fig.~\ref{fig3}(a), the structure consists of three concentric regions: a topologically nontrivial PC square of $6 \times 6$ unit cells, surrounded by a DZIM shell of 6-unit-cell thickness, which is further enclosed by a topologically trivial PC shell with 6 layers of unit cell. The calculated spectrum of this composite structure is shown in Fig.~\ref{fig3}(a). Without the DZIM layer, four nearly degenerate 0D corner states and twelve 1D edge states emerge at the closed interface between the topologically trivial and nontrivial PC (see Supplementary Materials for details), in accordance with bulk-edge correspondence. Interestingly, the insertion of the DZIM layer enforces a uniform electric field distribution along the closed interface, thereby suppressing three out-of-phase corner states and all 1D edge states. Moreover, the remaining in-phase corner state becomes distorted and spatially extended into the DZIM, resembling a hybridization with an in-phase 1D edge state. Consequently, its frequency deviates from that of the unperturbed corner state, falling at 5.3 GHz between the original corner state and the original edge state. Figure~\ref{fig3}(b) shows the spectrum when the innermost square region is replaced with a topologically trivial PC. In this case, the corner state within the band gap disappears, confirming that the emergence of the corner state is primarily determined by the topological difference between the inner and outer PC layers. Figure~\ref{fig3}(c) shows the normalized amplitude of the electric field of the spatially extended topological corner state within the entire three-layer concentric structure. Inside the DZIM layer, the electric field is uniform. In the topologically nontrivial region, the uniform field condition of the DZIM extends the in-phase topological corner state to the entire boundary of the nontrivial PC.

Using an effective parameter method based on field averaging at the boundary of eigenstates \cite{xu2021three,PhysRevLett.124.074501}, we find that the critical configuration between the topologically trivial and nontrivial PC, i.e., $L=5 ~\mathrm{mm}$, shows nearly vanishing effective permittivity and permeability around the frequency of the $X$ point in band structure (see Supplementary Materials). In view of this, we select the PC with $L=5 ~\mathrm{mm}$ as the filling. Figure ~\ref{fig3}(d) is the electric field distribution when the ideal DZIM in Fig. 3(c) is replaced by the PC with $L=5 ~\mathrm{mm}$. The essential characteristics of the DZIM and the spatially extended corner state are well preserved, making it possible to experimentally realize this phenomenon.

Furthermore, the spatial extension of the topological corner state remains robust even when the geometry of the DZIM shell changes. As shown in Fig.~\ref{fig3}(e), the four corners of the DZIM shell are modified by adding rectangular protrusions with an area equivalent to $2 \times 3$ unit cells. The in-phase topological corner state still extends into the deformed DZIM region, and its eigenfrequency remains nearly unchanged. Figure~\ref{fig3}(f) presents the electric field distribution when the ideal DZIM in Fig. 3(e) is replaced by a PC with $L = 5~\mathrm{mm}$. The spatial extension of the corner state remains, indicating that our strategy transcends conventional dimensional constraints of topological interfaces, enabling flexible geometric control of topological states, which is rarely discussed in previous studies.

\begin{figure}[b]
\centering
\includegraphics[width=1\columnwidth]{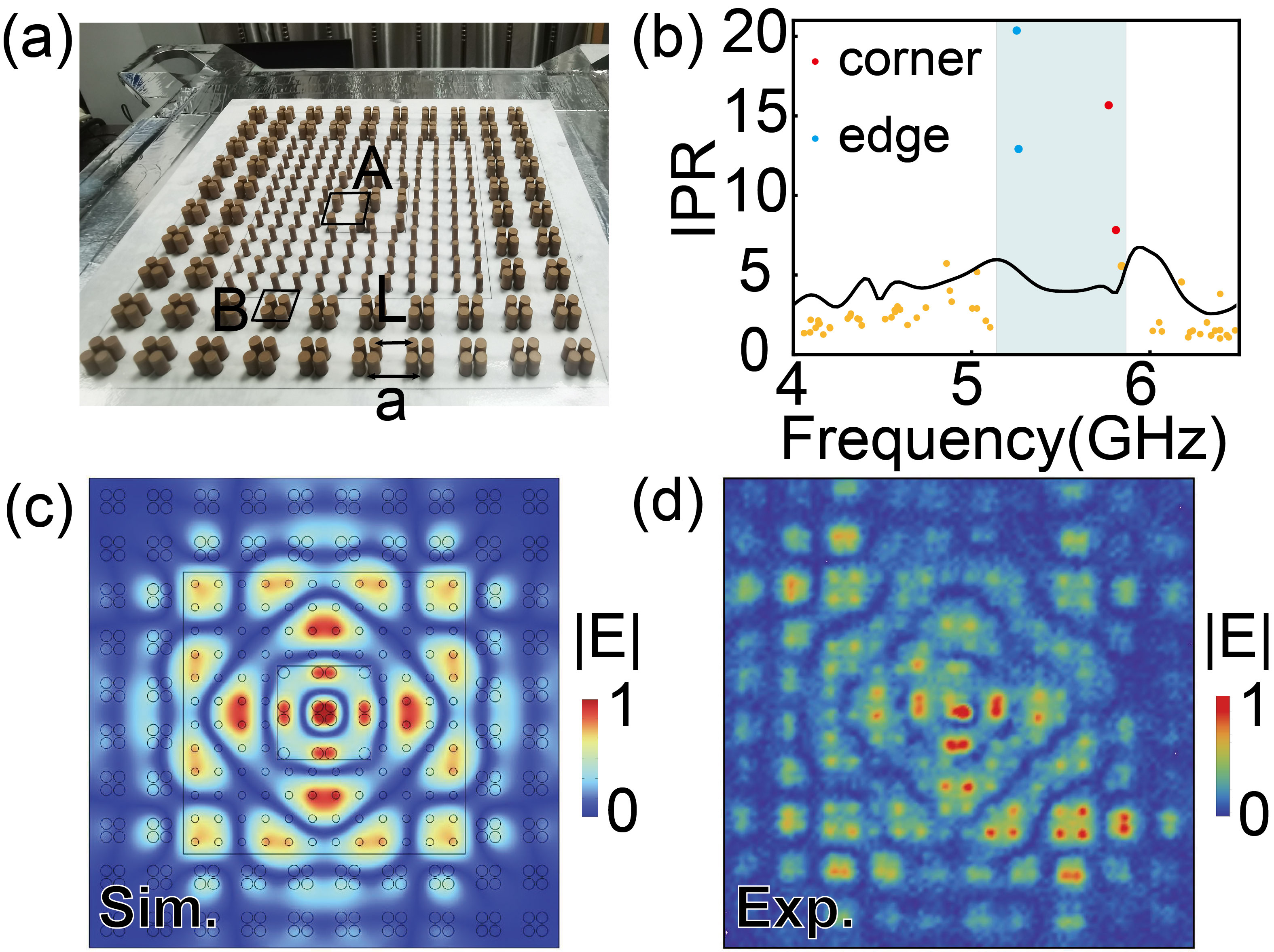}

\caption{Experimental realization of spatially extended topological corner states. (a) The PC composed of resin, with 2 layers of unit cell in each of the topological trivial layer, nontrivial layer, and the effective DZIM layer. (b) The IPR with and without the effective DZIM layer. (c, d) The electric field distribution of the spatially extended corner state from the (c) simulation and (d) experiment, respectively. }
\label{fig4}
\end{figure}

To experimentally verify this intriguing spatial extension of topological states, we fabricate a three-layer concentric PC with dielectric rods made of polyphenyl ether resin. Due to the size of our scanning platform, each PC layer consists of only two unit cells, which differs from the simulation. Figure \ref{fig4}(a) is a photo of the PC assembled on the surface of a metallic plate. Compared to the simulation setup, a slight adjustment is made: due to the limited number of periods in the interlayer that serves as a DZIM, the radius of rods in this region was set to $1.6~\mathrm{mm}$ to maintain the double-zero-index property. The reliability of this adjustment is confirmed by examining whether the electric field of the topological state remains spatially uniform at the boundary of the interlayer. The radius of each rod in the remaining region is set to $2.4~\mathrm{mm}$. To measure the electric field distribution near the frequency of the bulk band gap, we place an out-of-plane line source at the center of the PC and mount them on a stepper motor (see Supplementary Materials). In Fig. \ref{fig4}(b), we present the inverse participation ratio (IPR) of the electric fields with and without the DZIM. The IPR is defined as:

\begin{equation}
    {\rm{IPR}} = \frac{\int_{\rm{PC}} |E(r)|^4 \, dr}{\left( \int_{\rm{PC}} |E(r)|^2 \, dr \right)^2} \cdot A
    \label{eq:ipr}
\end{equation}
where $|E(r)|$ represents the amplitude distribution of the electric field and $A$ is the area of PC. The discrete colored points represent the IPR calculated from the eigenfields of the PC in Fig. \ref{fig4}(a), while the solid curve corresponds to the IPR obtained from the electric field distributions excited by the line source. Before the introduction of DZIM, the IPRs of the topological states are about 16, indicating strong localization of electric fields. However, when the PC serving as the DZIM layer is inserted at the interface between the topologically trivial and nontrivial bulks, the IPR of the electric field distribution drops significantly to around 4. This substantial reduction in IPR indicates a significant decrease in field localization, consistent with the spatial extension of the corner state. Figures \ref{fig4}(c) and \ref{fig4}(d) are the simulated and experimentally measured amplitude distributions of the electric field at $5.70 ~\mathrm{GHz}$, respectively, both clearly demonstrating the spatially extended corner state enabled by the effective DZIM.


In conclusion, we have proposed and experimentally demonstrated a novel strategy to overcome the intrinsic spatial confinement of topological states by embedding DZIM into topological interfaces. Owing to the curl-free nature of DZIM, this approach enables the free spatial extension of in-phase topological edge states and corner states, while selectively suppressing their out-of-phase counterparts. As a result, the traditional bulk–edge correspondence is effectively redefined, providing a new degree of freedom for manipulating topological states in PC. Our findings reveal that topological states can be flexibly shaped within extended spatial domains, paving the way for compact, integrable, and reconfigurable topological photonic devices. Given the universality of the mechanism, the proposed concept may also be extended to other classical wave systems, such as acoustics and elasticity.

\section*{Acknowledgements}
This work is funded by National Key R\&D Program of China Grant No. 2020YFA0211400 (Yun Lai), National Natural Science Foundation of China grant No. 12474293 (Yun Lai), No. 12174188 (Yun Lai), Jiangsu Specially Appointed Professor Program (Changqing Xu), Special funds for postdoctoral overseas recruitment, Ministry of Education of China (Changqing Xu), Natural Science Foundation of Jiangsu Province Grant No. BK20233001 (Yun Lai), No. BK20240576 (Changqing Xu), and the Chinese Academy of Sciences project No. E4BA270100 (Ce Shang), No. E4Z127010F (Ce Shang), No. E4Z6270100 (Ce Shang), and No. E53327020D (Ce Shang). 
\bibliography{apssamp}

\end{document}